\documentclass{article}
\usepackage{arxiv}

\usepackage{multicol}
\usepackage{xcolor,colortbl}
\usepackage{multirow}

\usepackage[utf8]{inputenc} 
\usepackage[T1]{fontenc}    
\usepackage{hyperref}       
\usepackage{url}            
\usepackage{booktabs}       
\usepackage{amsfonts}       
\usepackage{nicefrac}       
\usepackage{microtype}      
\usepackage{lipsum}
\usepackage{graphicx}
\graphicspath{ {./images/} }
\usepackage{natbib} 
\usepackage{graphicx}
\usepackage{helvet}
\usepackage{hyperref}
\usepackage{amsmath} 
\usepackage{amssymb} 
\usepackage{orcidlink} 
\usepackage{subcaption}
\usepackage{makecell}

\usepackage{colortbl}

\title{Improving understanding and trust in AI: How users benefit from interval-based counterfactual explanations}

\author{
 Tabea E. R\"ober \\
  Amsterdam Business School \\
  University of Amsterdam, The Netherlands\\
     \And
 Paul Festor \\
 Department of Bioengineering \\
Imperial College London, UK\\
   \And
 Rob Goedhart \\
Amsterdam Business School \\
  University of Amsterdam, The Netherlands\\
  \And
 \c S. \.Ilker Birbil \\
  Amsterdam Business School \\
  University of Amsterdam, The Netherlands\\
  \And
  Aldo Faisal \\
  Department of Bioengineering, Imperial College London, UK \\
  Institute of Artificial and Human Intelligence, University of Bayreuth, Germany \\
}

\begin{document}
\maketitle
\begin{abstract}
Experimental user studies evaluating the effectiveness of different subtypes of post-hoc explanations for black-box models are largely nonexistent. Therefore, the aim of this study was to investigate and evaluate how different types of counterfactual explanations, namely single point explanations and interval-based explanations, affect both model understanding and (demonstrated) trust. We conducted an online user study using a within-subjects experimental design, where the experimental arms were (i) no explanation (control), (ii) feature importance scores, (iii) point counterfactual explanations, and (iv) interval counterfactual explanations. Our results clearly show the superiority of interval explanations over other tested explanation types in increasing both model understanding and demonstrated trust in the AI. We could not support findings of some previous studies showing an effect of point counterfactual explanations compared to the control group. Our results further highlight the role individual differences in, for example, cognitive style or personality, in explanation effectiveness.
\end{abstract}

\keywords{Counterfactual explanations, User study, Interpretable AI, Explainable AI}

\section*{INTRODUCTION}\label{sec:intro}
Even though artificial intelligence (AI) models have enjoyed growing attention over the last decades, their implementation in high-stakes practice is still slow. A common critique centers around the difficulty to rely on such models especially in critical environments due to the black-box character of most models, which ultimately hinders user acceptance. On top of that, legal guidelines like the GDPR \citep{GDPR2016a} and the European AI Act \citep{eu_ai_act_2024}, were put in place to ensure that AI-based decision processes are sufficiently transparent and intelligible, or in other words, \textit{explainable} (XAI). A prominent type of explanations for predictions made by black-box models is \textit{counterfactual explanations}, which show how changes in the feature space would change the model prediction. This way, users may be able to form a better understanding of the model's decision boundaries. 

Specifically, algorithms finding counterfactual explanations search for a data point that is close to the original sample in terms of feature space, but which lies on the other side of the decision boundary. In a classification setting, the counterfactual would receive a different predicted class, whereas in a regression setting we would set a threshold that the counterfactual's prediction should exceed \citep{molnar_book2022, Guidotti2024}. It has been argued that counterfactual explanations are intuitive for humans because their contrastive nature (\textit{`Why did the model predict class A and not class B?'}) resembles human reasoning patterns in daily life \citep{byrne-ijcai2019p876, MILLER20191}. To date, almost all counterfactual explanation methods generate a single counterfactual data point as explanation \citep[\textit{e.g.},][]{Russell.2019, Ustun.2019, Kanamori.2020, Kanamori.2020jn, Karimi.2019fy, Karimi.2020, Mothilal.2020, Mahajan.2019, maragno2022counterfactual}. The feature values of this counterfactual can then be used to formulate an explanation in an if-then manner, \textit{e.g.} \textit{`If $x_1$ was smaller by x amount, then the model would have predicted class B'}. Instead of providing only a single counterfactual, \citet{maragno2024} propose to generate a region of possible counterfactuals. This way, an explanation can be formulated that refers to ranges in feature values instead of single data points, potentially providing more insightful information about the model's decisions. Methods generating these counterfactual intervals, also known as robust recourse, return a lower and upper bound for each feature. It is believed that this can be effectively translated into more user-friendly explanations that may give more agency to the user. It is further believed that those explanations might instill a broader understanding of the model behavior. 

At this point, many efforts have been made to propose algorithms that generate counterfactual explanations to enhance transparency of the model's decision process and by that increasing user understanding, trust, and acceptance of the models. However, only a few works conduct user evaluations to verify their claims that their models provide human-understandable output. In fact, only 7\% of papers proposing methods to generate counterfactual explanations perform any type of user evaluation \citep{DELANEY2023103995}, and hardly any of the conducted evaluations build on insights from the social sciences \citep{rong2024}. 

\begin{figure}[h!]
    \centering
    \includegraphics[width=1\linewidth]{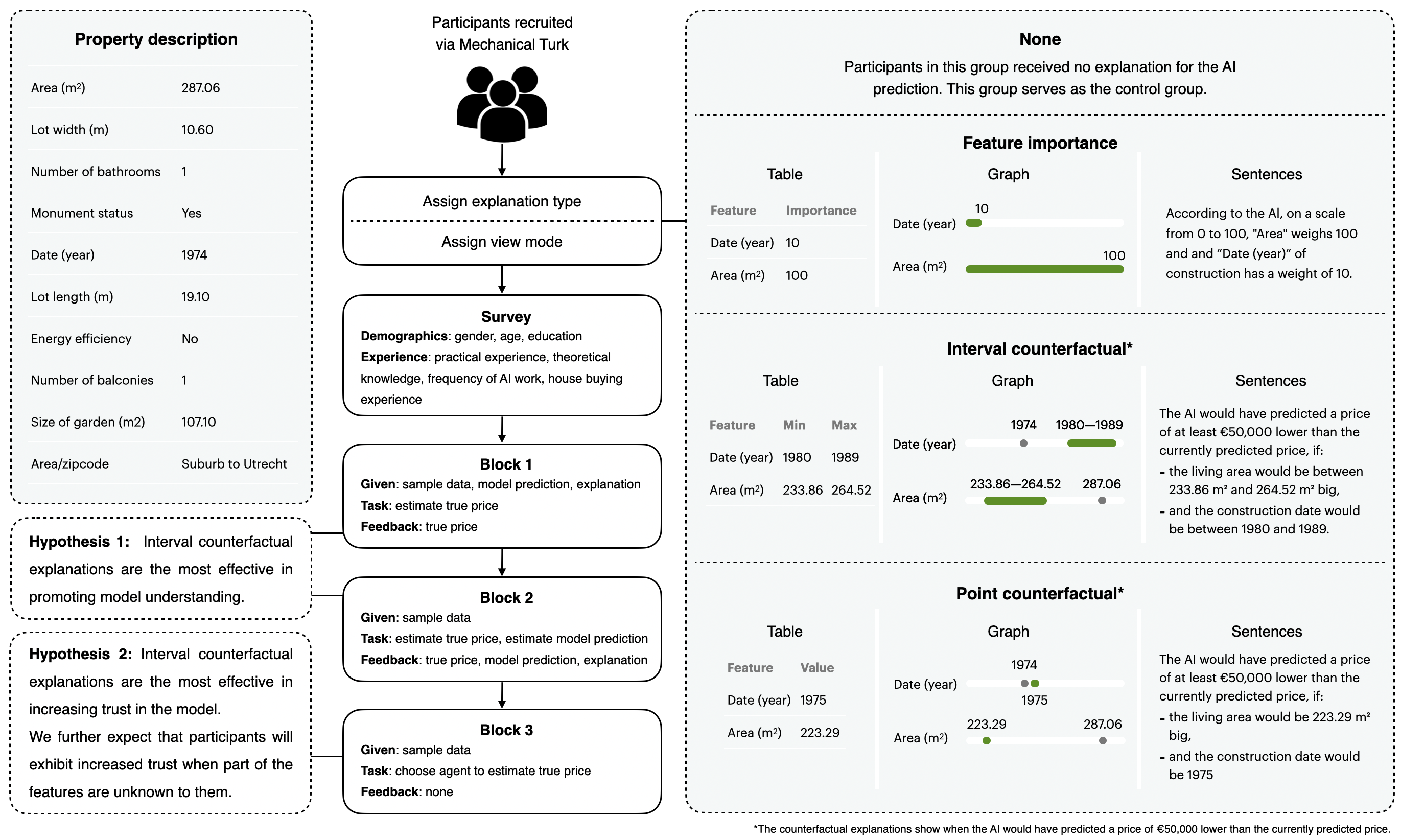}
    \caption{Schematic overview of the study concept and design. After filling out a survey to record demographics and experience level, each participant completes three blocks consisting of multiple trials each. Instructions within a block do not change. Participants are randomly assigned to one of the four explanation groups and one of the three different view modes.}
    \label{fig:study-concept}
\end{figure}

With this work, we aim to fill this validation gap by empirically studying \textit{how different types of counterfactual explanations impact user understanding and trust in the AI model}. In an online study environment, we ask users to estimate housing prices with the help of an AI model, and provide different explanations according to their assigned experimental arm. Participants receive either no explanation (control group), feature importance scores, point counterfactuals or interval counterfactuals (see Figure~\ref{fig:study-concept}). By having participants go through multiple blocks with different tasks each, we aim to address our research question. We expect that participants receiving any type of explanation show higher scores for understanding and trust. We hypothesize that interval counterfactual explanations are most effective in promoting understanding and in increasing reliance on the AI model. We further expect that reliance will generally increase when participants do not have full access to all feature information.

\section*{Results}\label{sec:results}

The experiment was successfully completed by 236 participants, of which 89 identify as female (38\%) and 147 as male (62\%). Participant ages ranged between 22 and 69 years, with a mean age of 34.28 (std=6.93) and a median age of 35. Participation took between 16 to 30 minutes and there was no significant difference in completion time of the experiment between groups (control: 1415.79s; feature importance scores: 1515.90s; point explanations: 1502.64s; interval explanations: 1485.02s; p=0.0699 by Kruskal-Wallis test). We refer to the supplementary material for detailed results of the models discussed in this section, such as regression coefficients and assumption checks.

\subsection*{Impact of explanation type on estimating house price}

\begin{figure}[h]
     \centering
     \begin{subfigure}[t]{0.48\textwidth}
         \centering
         \includegraphics[width=\textwidth]{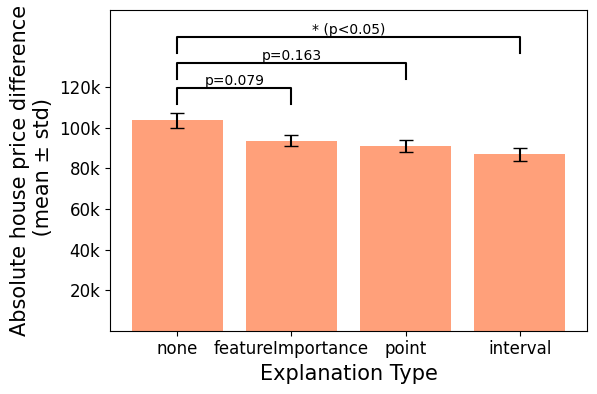}
         \caption{\small Adjusted group means of the house price differences.}
         \label{fig:p0-adjusted-group-means}
     \end{subfigure}
     \hfill
     \begin{subfigure}[t]{0.48\textwidth}
         \centering
         \includegraphics[width=\textwidth]{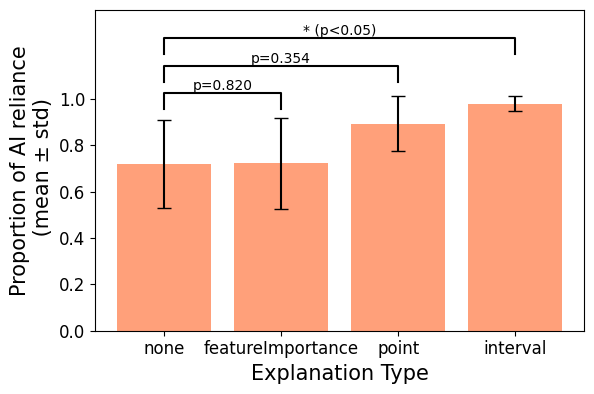}
         \caption{\small Adjusted group means of the proportion of AI reliance for trials with all feature information available.}
         \label{fig:p2-adjusted-group-means}
     \end{subfigure}
        \caption{Model-based group means by explanation type from two mixed-effects models with participant-level random intercepts, controlling for \textit{gender}, \textit{education}, and \textit{age}. Estimates are based on 100 bootstrapped samples per model. Error bars reflect bootstrapped standard deviations. Group means are estimated for the reference groups (\textit{female} and \textit{bachelors}) and the mean \textit{age} observed in the sample.}
        \label{fig:main-results}
\end{figure}

In the first block of trials, participants are provided with the values of features describing a house (\textit{e.g.} surface area, number of bathrooms, energy efficiency, etc.; see Section~\nameref{sec:methods}), as well as the AI's prediction of the house price, and the corresponding explanation. Participants are then asked to give an estimate of the true house price given this information. The target variable is the absolute difference in price between their estimate and the true house price, on which we ran a mixed effects model with a random intercept for participants. Controlling for demographics (\textit{gender}, \textit{education}, \textit{age}), participants receiving interval explanations score significantly better than the control group ($\beta_{interval}=-0.164, p_{interval}<0.05$). Other explanation groups did not outperform the control group ($\beta_{featureImportance}=-0.097, p_{featureImportance}=0.163$; $\beta_{point}=-0.123, p_{point}=0.079$), however we do see a trend in the hypothesized direction. Estimated group means for absolute house price difference adjusted for \textit{gender}, \textit{education}, and \textit{age} are displayed in Figure~\ref{fig:p0-adjusted-group-means}. Further controlling for other potential confounding variables related to experience (\textit{practical AI experience}, \textit{theoretical knowledge of AI}, \textit{frequency of work with AI}, \textit{recent house buying experience}) or study design (\textit{explanation view mode}), the observed effect stays the same ($\beta_{interval}=-0.159, p_{interval}<0.05$).

Trials in the second block differ from those in the first block as participants only receive the house information and are then asked to indicate both their estimate of the true house price and what they think the AI would predict. This setup is intended to gauge whether participants pick up on the model's behavior throughout the first block, which is related to model understanding. Two mixed effects models with random intercepts for participants were fit on the absolute difference in house prices (estimate and true price) and the absolute difference in AI prediction (estimate and true prediction). No significant difference between experimental arms was observed.

\subsection*{Impact of explanation type on reliance}

In the last block, participants needed to decide whether the AI should predict the house price or whether they want to rely on their own estimate instead. A logistic regression model with random intercepts for participants was fit to the data, with the target variable indicating whether participants relied on the AI or not. Participants in the interval explanation group chose the AI as agent significantly more often than participants from the control group ($\beta=3.989$, $p<0.01$). This effect remained also when controlling for demographics (\textit{gender}, \textit{education}, \textit{age}), experience (\textit{practical AI experience}, \textit{theoretical knowledge of AI}, \textit{frequency of work with AI}, \textit{recent house buying experience}) or study design (\textit{explanation view mode}). Again, the other two treatment groups show no significant difference to the control group, \textit{i.e.}, participants receiving feature importance scores or point counterfactuals as explanation are equally likely to rely on the AI as those receiving no explanation at all. In Figure~\ref{fig:p2-adjusted-group-means} we display the estimated group means adjusted for \textit{gender}, \textit{education}, and \textit{age}.

In half of the trials in this block, three out of 10 features were masked. In other words, participants did not receive the complete house information. In this case, there was no significant difference between any of the treatment groups and the control group.

\subsection*{Effect of AI reliance on house price prediction accuracy}

\begin{figure}[h!]
     \centering
     \begin{subfigure}[b]{\textwidth}
         \centering
         \includegraphics[width=\textwidth]{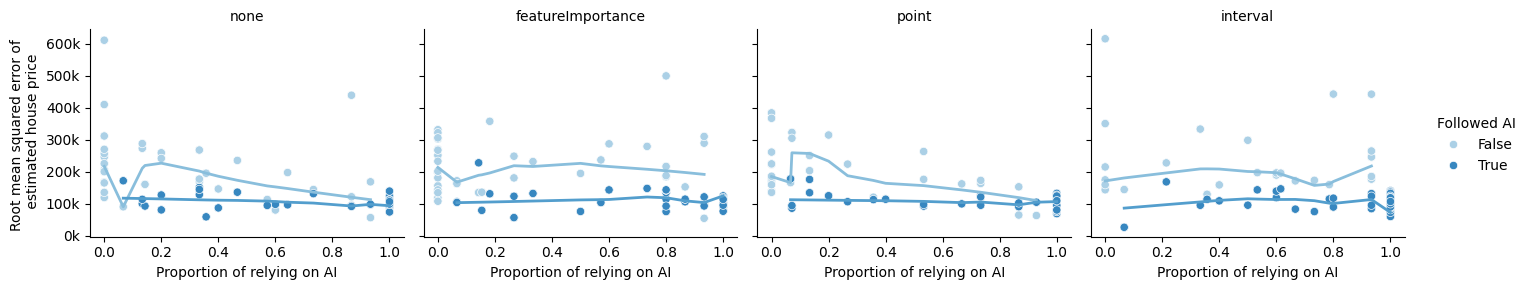}
         \caption{\small RMSE of estimated price by following behavior and explanation type for trials with no features hidden.}
         \label{fig:p2-rmse-hiddenFalse}
     \end{subfigure}
     \hfill
     \begin{subfigure}[b]{\textwidth}
         \centering
         \includegraphics[width=\textwidth]{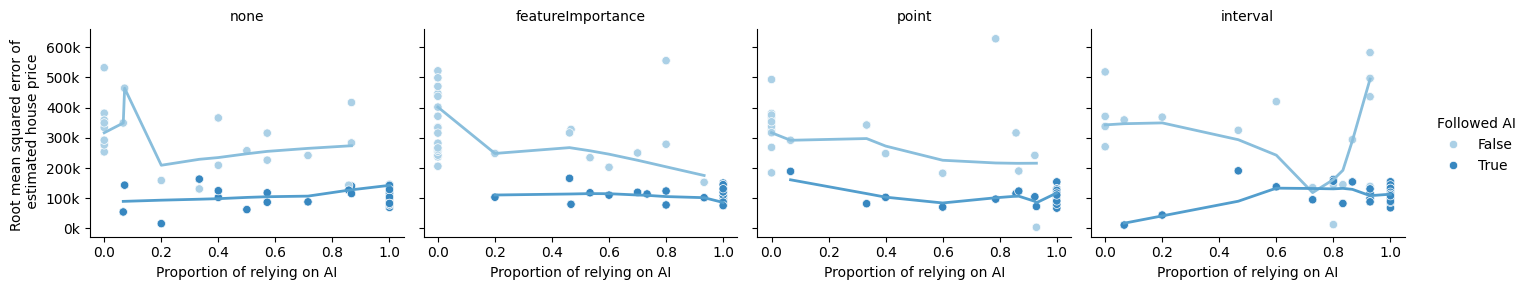}
         \caption{\small RMSE of estimated price by following behavior and explanation type for trials with features hidden.}
         \label{fig:p2-rmse-hiddenTrue}
     \end{subfigure}
        \caption{Root mean squared error (RMSE) by following behavior vs reliance on AI proportion, for trials where no features are hidden and where three out of 10 features are hidden. Each user is represented by two dots (unless the proportion is 0 or 1). A LOWESS curve is shown to visualize trends in prediction error.}
        \label{fig:p2-rmse}
\end{figure}

If participants chose to rely on themselves to make the prediction of the house price, they were prompted to provide their estimate. In Figure~\ref{fig:p2-rmse} we display the root mean squared error (RMSE) per user by following behavior and by proportion of reliance on the AI, for trials where features were shown (Figure~\ref{fig:p2-rmse-hiddenFalse}) and hidden (Figure~\ref{fig:p2-rmse-hiddenTrue}). We plot a locally weighted scatterplot smoothing (LOWESS) curve for trials where participants followed the AI and did not follow it to visualize trends in the prediction error as a function of participants' reliance on AI. 
Across groups, we see that participants tend to make a larger error when relying on themselves than when following the AI. 
We further see that errors tend to be larger in trials where features were hidden (Figure~\ref{fig:p2-rmse-hiddenTrue}).

\subsection*{The role of gender for the effectiveness of explanations}

\begin{figure}[h!]
     \centering
     \begin{subfigure}[b]{0.49\textwidth}
         \centering
         \includegraphics[width=\textwidth]{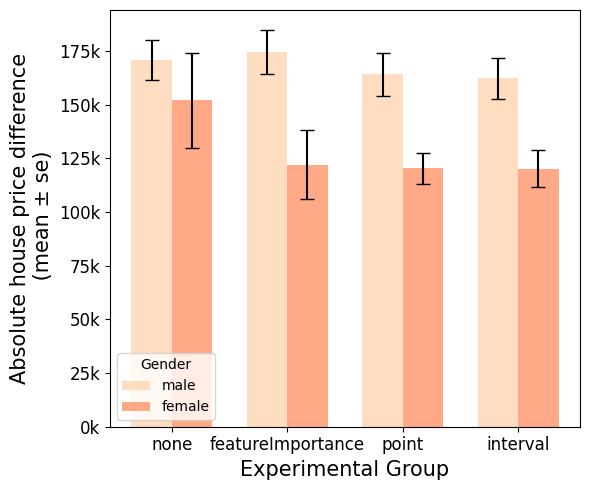}
         \caption{\small Observed group means of the house price differences averaged by user (block 1).}
         \label{fig:p0-mean-scores-gender}
     \end{subfigure}
     \hfill
     \begin{subfigure}[b]{0.43\textwidth}
         \centering
         \includegraphics[width=\textwidth]{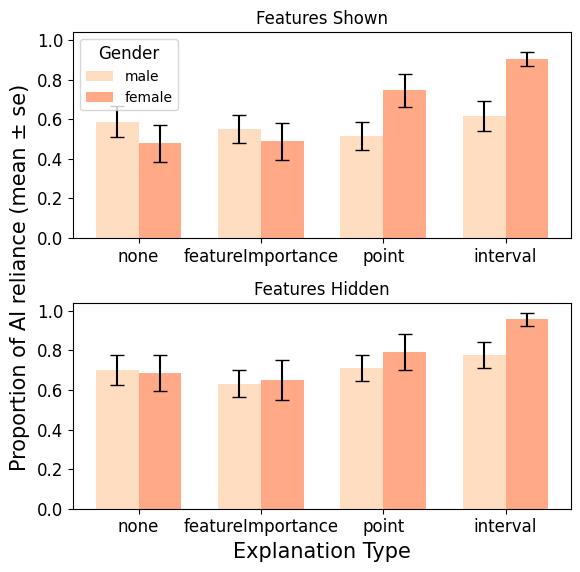}
         \caption{\small Observed group means of the proportion of AI reliance by gender.}
         \label{fig:p2-mean-scores-gender}
     \end{subfigure}
        \caption{Observed group means averaged per user by explanation type and gender.}
        \label{fig:results-gender}
\end{figure}

Across models, we observe a significant effect of gender on both model understanding and reliance behavior, with male participants consistently demonstrating lower levels of understanding (larger error) and less reliance on AI. In Figure~\ref{fig:results-gender} we plot mean scores observed in the data for some of the dependent variables by gender. Specifically, Figure~\ref{fig:p0-mean-scores-gender} shows the absolute differences in house price from block 1 trials averaged per user by explanation type and gender, and Figure~\ref{fig:p2-mean-scores-gender} shows the proportion of relying on the AI (block 3) by explanation type and gender observed in the data, for trials where all feature information was shown and when part of them where masked. This effect remained significant even after controlling for covariates, including those that gender correlates moderately with ($\rho_{gender,education} = 0.37$, $\rho_{gender, age}=0.22$, $\rho_{gender,submissionTime}=-0.24$). This suggests that the gender difference is not fully explained by these factors. In our supplementary material, we have added a thorough analysis of gender and other background variables we recorded to rule out alternative explanations for this effect. Although a gender effect was not a central focus of our experimental setup, its persistent association with the outcome variables raises important questions about potential gender-based differences in the effectiveness of counterfactual explanations.

\section*{Discussion}\label{sec:discussion}

\subsection*{Main findings}

This study yields multiple interesting and valuable findings for AI-assisted decision making that should be carefully considered when designing AI-based recommendation systems and that give rise to interesting future research avenues. First, we observe that interval counterfactual explanations outperform explanations based on feature importance scores or point counterfactuals with respect to model understanding (Figure~\ref{fig:p0-adjusted-group-means}).
While we expected that receiving any type of explanation would lead to significantly different scores than receiving no explanation, this was only the case for interval explanations. Moreover, although trends aligned with our hypotheses, we found no statistically significant differences between explanation types.
Second, another aspect of model understanding evolves around how good participants are in mimicking the AI's prediction, however no significant differences were observed between experimental groups in this regard. This suggests that while interval explanations help participants make a better judgment about the true price when seeing both the AI prediction and the explanation, it does not lead to significantly improved replicating performance.
Third, participants relied significantly more often on the AI system than on themselves in the third block when receiving interval explanations compared to the control group, suggesting that interval explanations instill a higher sense of trust in participants (Figure~\ref{fig:p2-adjusted-group-means}). This effect is only present in scenarios where participants receive full information about the house. In cases where information was partially masked, participants in all groups chose to rely on the AI in significantly more than 50\% of the trials. However, there is no significant difference between any of the treatment groups. Fourth, gender has a significant effect on performance and AI reliance, which are proxies for model understanding and trust, respectively (Figure~\ref{fig:results-gender}). On average, male participants showed larger errors, which indicates lower model understanding, and they relied less often on the AI, which may indicate that they trust the AI less. 

\subsection*{Model understanding}
Our main contribution is the empirical evaluation of interval counterfactual explanations, a recently introduced form of counterfactual explanation that extends beyond the traditional point counterfactuals. Several previous studies, focusing on point counterfactual explanations, have shown that those can positively impact users understanding of AI-based systems, both in terms of objectively measured task performance \citep[\textit{e.g.},][]{Kuhl2023, WIJEKOON2023110830, Mertes2022, warren2022features, Silva2023} and in terms of self-reported measures of perceived usefulness and satisfaction \citep{Kuhl2023, RIVEIRO2021103507, STEPIN2022379}. These findings have fostered the widespread assumption that point counterfactual explanations pose an effective means to promote model understanding. However, while subjective measurements seem to consistently support this assumption, an objective improvement in model understanding induced by point counterfactual explanations could not be confirmed by other studies \citep{RIVEIRO2021103507, VANDERWAA2021103404, warren2022features}.
In this study, we demonstrated that while point counterfactuals may offer some insights into model understanding, their effects were not statistically significant. In contrast, our findings clearly indicate that interval counterfactual explanations significantly enhance model understanding compared to the control group. This underscores that while point counterfactuals may seem informative, their effectiveness varies across different domains and tasks. Our results firmly establish the superiority of interval-based explanations, as they provide users with a range of feature values rather than a single static data point. This approach yields substantial benefits in both understanding and trust, particularly when complete input information is available.

This finding aligns with previous work that explored how interval-based explanation explain (large) prediction errors rather than the prediction itself \citep{lucic2020}. In this context, the explanations return feature ranges for the most important features for which the resulting prediction would have a high confidence. While average accuracy for task performance centers around 81\% \citep{lucic2020}, this study lacks a comparison to the control group and hence no conclusion can be drawn whether those explanations result in a significant improvement. Therefore, our study design and analysis of explanation types compared to a control group with no explanations provides a stronger foundation for the claim that interval counterfactuals support deeper understanding of the underlying AI. Another related finding of previous work underscores that humans generate broader, more prototypical counterfactuals when asked to create their own counterfactuals for images \citep{DELANEY2023103995}, which suggests that users may benefit from more generalizable insights into model behavior rather than static point explanations. Our study extends these findings to structured tabular data.

Interestingly, however, our results do not show a significant difference between explanation types in participants' ability to mimic the AI's behavior (block 2). This suggests that, even though explanations help users in the decision-making process when they complement the AI prediction, there is limited transfer of understanding. These results reinforce the idea that explanations may aid interpretation of individual predictions, however they do not necessarily enable users to accurately replicate the AI's decisions.

\subsection*{Demonstrated trust}
Beyond model understanding, it is believed that model understanding promotes trust in AI systems. However, empirical findings are mixed: while some studies find an effect of counterfactual explanations on trust \citep[\textit{e.g.},][]{Kuhl2023, warren2022features}, this could not be confirmed in other studies \cite[\textit{e.g.},][]{lucic2020, Silva2023}. Typically, trust in empiricial user studies is measured through self-report scales \cite[\textit{e.g.},][]{Mertes2022, warren2022features, lucic2020}. 
However, self-report scales are increasingly criticized due to, for example, their unreliability, susceptibility to biases, and misinterpretation of Likert items, which has been repeatedly raised in recent work around trust in AI models \cite[\textit{e.g.},][]{miller2022we, Jacovi2021}. 
Therefore, our study focuses on assessing \textit{demonstrated} trust, which can be measured through reliance behavior \citep{miller2016behavioral}. Our results show that only interval counterfactual explanations significantly increased reliance compared to the control group. Again, while trends in the point counterfactual group are in line with previous findings stating a positive effect of counterfactual explanations on trust, those effects were not statistically significant. Notably, our study adds further nuance by showing that the found effect is conditional on information availability. That is, in scenarios where participants had full access to feature information, interval explanations promote demonstrated trust over no explanations. However, this effect vanishes when key features are hidden. In this case, participants from all groups, including the control group, rely on the AI in the majority of scenarios and at similar rates. This suggests that uncertainty or information availability plays a key role in explanation effectiveness.

\subsection*{The role of gender in explanation effectiveness}
The observed gender effect in our study deserves closer attention. Although our study was not designed specifically to identify differences caused by gender, we found that male participants, on average, exhibited lower model understanding and lower levels of reliance on the AI. Importantly, we were unable to identify any other recorded variable that accounts for this difference. This raises the possibility that unmeasured factors, such as personality traits, confidence levels, AI-related anxiety, or attitudes towards AI, may play a key role in explanation effectiveness. In a recent study, \citet{russo2025} show that women exhibit higher levels of AI anxiety, less positive attitudes toward AI, and lower self-reported knowledge compared to men. These characteristics may influence how users interact with AI-based systems. For example, users with higher AI-anxiety and more negative attitudes toward AI may act more cautiously, by that engaging in more attentive processing of given explanations, which could enhance model understanding. On the other hand, lower levels of concern may lead users to superficial engagement with explanations or overconfidence in their own judgment. Further building on findings from \citet{russo2025}, we would expect men to show higher reliance on the AI system due to their, on average, greater reported trust and more favorable attitude towards AI. However, our sample shows the opposite, that is women relied significantly more often on the AI. A possible interpretation is that, as shown in previous studies, model understanding fosters trust \citep[\textit{e.g.},][]{lim2009,Kuhl2023, warren2022features}, and we found in the first part of our study that women formed a better understanding of the model, which could have increased their trust and reliance on the AI.

On a broader level, these findings contribute to a growing body of literature on gender disparities and gender bias in XAI. For example, \citet{dhaini2025} demonstrate that popular explanation methods display significant gender-disparities with regard to robustness and faithfulness. \citet{JussupowMezaMartinezMaedche2021_1000139962} study how gender-biased explanations affect trust, and report that users with low awareness of societal gender stereotypes are more likely to trust gender-biased systems, likely because it is nonetheless more transparent than a system without any explanations. Extending discussions of gender bias to fairness, \citet{dodge2019} test how different explanations types support participants to detect case-specific unfairness. Together, these studies highlight the need to incorporate fairness considerations into the design and evaluation of explainability methods.

\subsection*{Limitations and future research}
Our findings should be interpreted in light of several limitations. First, the chosen context and task were relatively complex compared to other study designs. While this was an intentional choice to mimic real-world applications, it adds a layer of complexity to the study design. In an online study environment, participants may lack intrinsic motivation to commit to the study if it requires increased mental load. Second, the completion time for the study was relatively long, as we aimed to study both model understanding and demonstrated trust, which may increase the risk of superficial and unfaithful responses. Although we implemented quality control measures (\textit{e.g.}, completion thresholds) to mitigate this issue, these are not foolproof. Third, participants were recruited via an online platform, which may limit the generalizability of our results to real-world applications. Fourth, we constrained our study design to focus on different explanation types without manipulating explanation quality or model accuracy, which may interact with users' trust and understanding. Lastly, our results show notable gender differences that go beyond differences in other background variables such as education or AI knowledge. As we did not measure other potential moderating factors, such as AI anxiety or personality traits, we are unable to pinpoint the underlying mechanisms for the observed difference.

Interesting future research avenues include exploring how explanation quality and model accuracy affect and interact with users' demonstrated trust in addition to perceived trust, with a special focus on calibrated trust. While our study meets the three out of four identified requirements to measure calibrated trust \citep{miller2022we}, a study set-up that introduces controlled variations in model accuracy is necessary to disentangle whether exhibited trust in treatment groups was warranted. In addition, explanation effectiveness with regard to individual differences in, for example, personality, cognitive styles, or attitudes, should be investigated further to understand how inclusive explanations should be designed. 
Besides, we focus our work on single point explanations and interval explanations. Other forms of counterfactual explanations, such as sets of multiple single ones \citep[\textit{e.g.},][]{Mothilal.2020, Russell.2019, Kanamori.2020jn} or probabilistic counterfactuals \citep{pawelczyk2022probabilistically}, could be investigated too. However, some of these methods do not guarantee that generated counterfactuals are valid \citep{Mothilal.2020, pawelczyk2022probabilistically}, which could raise questions about their usability in practice as well as in a study similar to ours. It would further be valuable to replicate our study with a different context to validate findings, and extending the evaluation of interval explanations to domains with high-stakes decisions (\textit{e.g.}, healthcare, finance) could shed light on their utility in real-world applications. 

\subsection*{Conclusion}
To conclude, the goal of this study was to evaluate the effectiveness of point and interval counterfactual explanations on users' understanding and trust in AI-assisted decision processes. While prior work has found support for point counterfactual explanations, our findings did not confirm these effects. Instead, only interval explanations showed statistically significant improvements in both model understanding and demonstrated trust over the control group. Our study highlights the need for future research to further study interval counterfactual explanations with real-world applications and to investigate the role of individual differences in their effectiveness.


\vspace{0.9cm}
\section*{Methods}\label{sec:methods}

\subsection*{Experimental procedure and design}\label{sec:procedure}

Participants were recruited via Amazon Mechanical Turk (AMT)\footnote{\url{https://requester.mturk.com/create/projects/new}}. After signing up, participants were re-directed to a web server hosting the study. Once there, users were first provided full information about the purpose of the study, the procedure, the data collected and how their data will be handled, as well as the contact details of the primary investigator. Furthermore, participants were informed about their options to withdraw from the study, which they could do at any time and without giving a reason. The study began once participants indicated their agreement by selecting the relevant box and confirmed via button press. The study itself consisted of two main parts: a survey and the experiment. 

\paragraph{\textbf{Survey}}\label{sec:survey}
The purpose of the survey was to record relevant background information and demographics of the participants to acquire an impression of the sample and to be able to include potential confounding variables in the analysis later on. Demographics recorded included the gender they identify with, age, and level of education. Then, participants were asked to rate their theoretical knowledge about AI and their practical experience with AI on a scale from zero to 10, where zero means `no experience' and 10 refers to `expert'. In addition, we asked them to indicate the frequency at which they are working with AI (daily/weekly/monthly/less than monthly) and whether or not they had bought or actively considered buying a house in the past five years (yes/no), which is tied to the dataset and the task used in this study (see Section~\nameref{sec:materials}). Lastly, as the study included prices and area metrics, we asked them to indicate their currency preference (US Dollars (\$), British Pound (£), or Euro (€)) and their area metric preference ($m^2$ or $ft^2$). After filling in these questions, participants received information about the remainder of the study, specifically about the structure and the dataset that will be utilized in the experiment. 

\paragraph{\textbf{Experimental design}}\label{sec:design}

A schematic overview of the experimental design is displayed in Figure~\ref{fig:study-concept}. As our primary interest is the effect of different types of explanations on model understanding and trust, we introduce four experimental conditions, where each participant will be randomly assigned to one of them. The conditions are \textit{no explanation}, \textit{feature importance scores}, \textit{point counterfactual explanations}, and \textit{interval counterfactual explanations}. To exclude that differences in effects are due to the style of presentation of the explanation, we further introduced three different ways of displaying explanations: natural language sentences, tabular format, and a visual display. Again, participants are randomly allocated to one of these display options. The experiment is split into three phases that each participant goes through in the same order. Each phase consists of multiple trials that require the participant to perform a specific task. Tasks within a phase do not differ in terms of the instructions. Therefore, participants are presented with the instructions at the start of each phase. However, they could always revisit them at any time during the experiment. 

In the first block, participants received information about a single sample from the dataset, \textit{i.e.}, they could view the feature values of that sample. Furthermore, they saw the model prediction and the explanation corresponding to the condition they were assigned to. Naturally, the `no explanation'-condition only saw the model prediction without any explanation. Then, the task entailed providing an estimate of the true target value for this sample based on all available information. After submitting their estimate of the true target value, participants received feedback in terms of the true target value (the `ground truth'). A total of 20 trials were administered in this block.

In the following block, participants were again shown feature values of a single sample, however this time without model prediction and explanation. Participants were asked to provide two estimates: one for the ground truth and one for the AI prediction. Once submitted, they received feedback including the true target value (ground truth), the actual AI prediction, and the explanation for this prediction. Again, this block consisted of 20 trials.

The last block was intended to measure demonstrated trust, and again participants received the feature values of a single sample at a time. In this phase, participants were asked to choose an agent -- either the AI or themselves -- to estimate the house price. If they chose to predict it themselves, they were prompted to provide their estimate. Participants did not receive any feedback in this phase, and this was repeated for 30 trials. We further masked 30\% of the features in half of the trials to simulate a situation of limited knowledge about the scenario. To incentivize participants to choose wisely, we simulated a risk by showing participant their performance score. Every participant started off with a score of 100, and after each trial the score went down proportionally to the size of the gap between the predicted target value and the ground truth value. Participants were informed that they will have a chance to win a gift card of £50, and their chance is relative to their score. With this potential gift, participants were encouraged to carefully consider their choice of agent.

\subsection*{Materials and implementation}\label{sec:materials}

Throughout the entire experiment, we utilized the same dataset on which we trained a black-box model. Subsequently the three explanation methods were applied to the trained model. The details are given below.

\paragraph{\textbf{Dataset}}\label{sec:dataset}

The dataset that was chosen to model the task is a regression dataset containing information about houses and their sales prices in the Netherlands. The dataset is publicly available and was made available by Utrecht University of Applied Science\footnote{\url{https://ictinstitute.nl/utrecht-housing-dataset/}} in 2022. While synthetic, the dataset was based on real data of the Dutch housing market. The dataset consists of 2000 rows and 13 features that are described in Table~\ref{tab:dataset}. The target is \textit{retailvalue}, which is the market value of a house rounded to the nearest €1000.00.

{\renewcommand{\arraystretch}{1.5}
\begin{table}[h!]
\centering
\caption{Overview of features included in the housing dataset utilized in this study}
\label{tab:dataset}
\resizebox{0.9\textwidth}{!}{
\begin{tabular}{lll}
\hline
\textbf{Feature}  & \textbf{Description}       & \textbf{Type}       \\
\hline
\rowcolor{gray!10}  zipcode & zipcode corresponding to the area the house is located in & categorical (4) \\ 
lot-len & the length in meters of the plot of land the house is built on & float \\
\rowcolor{gray!10} lot-width & the width in meters of the plot of land the house is built on & float \\
lot-area & the total area of the plot of land the house is built on & float \\
\rowcolor{gray!10} house-area & the living area of the house in square meters & float \\
garden-size & the size of the garden in square meters & float \\
\rowcolor{gray!10} balcony & the number of balconies the house has & int / categorical (1/2/3) \\
x-coor & the x-coordinate describing the location of the house & int \\
\rowcolor{gray!10} y-coor & the y-coordinate describing the location of the house & int \\
buildyear & the year that the house was built & int \\
\rowcolor{gray!10} bathrooms & the number of bathrooms the house has & int / categorical (1/2/3) \\
energy-eff & energy efficient house & binary (0/1) \\
\rowcolor{gray!10} monument & house with monumental value & binary (0/1) \\
\hline
\end{tabular}
}
\end{table}}

\paragraph{\textbf{AI model}}
The model that was trained on this dataset and acted as `black-box' model for the experiment was a random forest regressor. The target is a continuous feature representing the market value of a house (\textit{retailvalue}). As input features, we included area of a house (\textit{zipcode}), length and width of the plot (\textit{lot-len} and \textit{lot-width}), living area of the house (\textit{house-area}), garden size (\textit{garden-size}), the number of balconies and number of bathrooms (\textit{balcony}, \textit{bathrooms}), the year of construction (\textit{buildyear}), whether or not the house was energy efficient and of monumental value (\textit{energy-eff}, \textit{monument}). 
The area of the plot was removed as this amounts to the product of length and width of the plot, and x- and y-coordinates were also removed as those were expected to bear no semantic meaning to participants. While 10 features is quite substantial number of features, we opted for not reducing the number further because it is expected that many real world applications with tabular data structures will have a similar or an even higher number of features. At no point were participants informed about any technical details relating to the model, but instead all instructions simply referred to `the AI' to elucidate the impression of a very complex model. The train and test $R^2$ scores were 0.78 and 0.77, respectively, however this was not communicated to the participants. We did not perform substantial model tuning as we aimed to utilize a model with a decent but not outstanding performance such that participants will not be inclined to rely on the model in 100\% of the trials in the last phase of the experiment.

\paragraph{\textbf{Post-hoc explanation methods}}
For the feature importance scores we opted to apply SHAP \citep{Lundberg-SHAP} as it one of the most popular and commonly used feature importance methods for tabular data. SHAP, short for SHapley Additive exPlanations, originates in cooperative game theory and estimates each feature's contribution to the prediction, by that breaking down a model's prediction. For point counterfactual explanations we applied CE-OCL \citep{maragno2022counterfactual}, an optimization-based approach that finds the closest counterfactual by modeling the problem as an optimization with constraint learning problem \citep{FAJEMISIN20241}. Lastly, we relied on RCE proposed by \citet{maragno2022counterfactual} to generate interval counterfactual explanations. This approach is based on robust optimization techniques and works well for many common machine learning models including tree ensembles. We chose a radius of $\rho=0.05$ for the uncertainty set for RCE. Since this is applied to scaled data, the resulting box has a total width of 0.1 in each dimension. This corresponds to 10\% of the full range of the scaled data. For both counterfactual methods we applied the following settings to ensure both methods have the same constraints and return realistic explanations. The threshold chosen was a retail value of 50.000 less than the true retail value of a house. In other words, each explanation showed the changes required in input features to have the model predict a retail value of 50.000 less than the actually predicted value. Furthermore, we fixed features \textit{zipcode}, \textit{energy-eff}, and \textit{monument} to be immutable, and defined a minimum value for \textit{bathrooms} (1) and \textit{house-area} (30$m^2$). Lastly, the construction year was constrained to be at most 2024 as the study was run in early 2025.

\subsection*{Participants}\label{sec:participants}

This study was conducted online and participants were recruited via AMT. Inclusion criteria for participation was a minimum age of 18 and being fluent in the English language. No prior knowledge about the subject matter was required, however some background information was recorded to use as potential confounding variables in the analysis (see details in Section \nameref{sec:analysis}). Participation was voluntary and all participants received a compensation of £1.50 for successful completion of the study. A power analysis was conducted to get an indication of the number of participants required. The final sample of valid participants, \textit{i.e.}, those that filled out the survey completely and according to our quality checks, consisted of 236 participants, of which 89 female (38\%) and 147 male (62\%). Participants were aged between 22 and 69 years, with a mean age of 34.28 (std=6.93). Regarding education level, 156 participants (66\%) held a bachelors degree or equivalent, 77 (33\%) a masters degree or equivalent, one had a PhD and three completed highschool as highest education. The study was approved by the Research Governance and Integrity Team (RGIT) of Imperial College London as well as by the Economics and Business Ethics Committee (EBEC) of the University of Amsterdam. Participants were informed in full about the research goal, study procedure, data analysis and their rights. All participants signed an provided informed consent prior to participation.

\begin{figure}
    \centering
    \includegraphics[width=\linewidth]{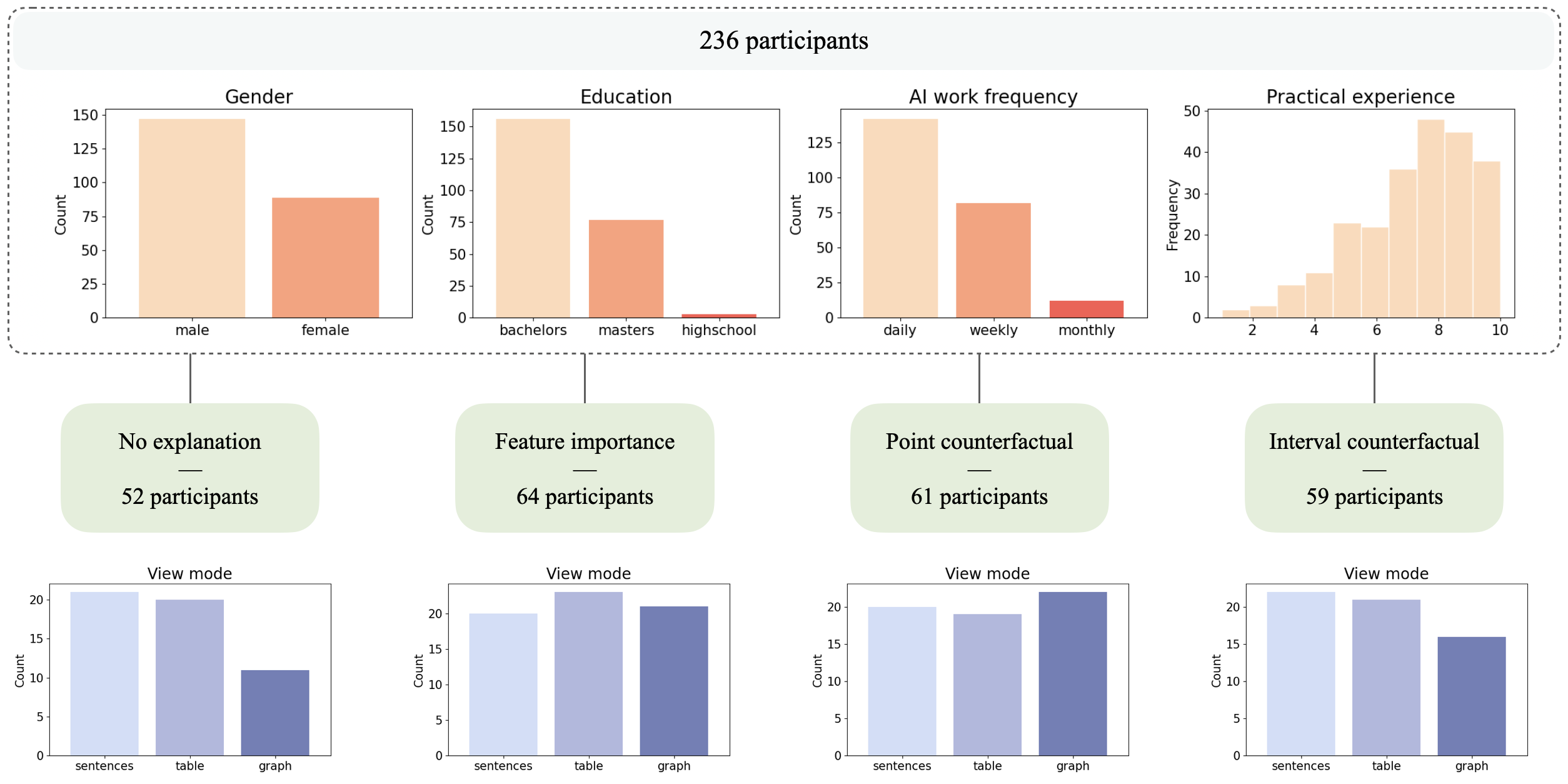}
    \caption{Distribution of participants over conditions}
    \label{fig:participant-dist}
\end{figure}

\subsection*{Analysis}\label{sec:analysis}

\paragraph{\textbf{Data preparation}}

Outliers in the variables of interest, that is the participants' estimated property value and estimated AI prediction, were removed using the Interquartile Range  (IQR) method, where data points outside $Q1 - 1.5\times IQR$ and $Q3 + 1.5\times IQR$ are considered extreme values. Here, $Q1$ is the median of the first quartile and $Q3$ is the median of the third quartile. For data collected in the first two blocks, a new variable indicating the absolute difference between the true retail price and participants' estimate is created. Likewise, for data from the second block, another variable representing the absolute difference between the AI's prediction and the participant's estimate of the AI's prediction was created. These serve as the dependent variables in the further analyses. All continuous (independent and dependent) variables were scaled and centered to aid interpretation of coefficients. Lastly, categories `PhD' and `masters' of education level were merged as well as `less\_than\_monthly' and `monthly' of AI work frequency were merged into one as there were not enough observations in `PhD' and `less\_than\_monthly' respectively.

\paragraph{\textbf{Assumption check}}

Our analysis relies on certain assumptions that were checked for each model. Normality of residuals was investigated mostly visually by inspecting the Q-Q-plot and histogram. Furthermore, skewness and kurtosis values were calculated and evaluated with common rule of thumbs to judge whether the distribution is sufficiently normal. That is, absolute skewness and kurtosis values $< 1.00$ were considered to be sufficient. A Shapiro-Wilk test for normality was not performed as results of this test are unreliable with $N>5000$. Homogeneity of variance was investigated with Levene's test for equal variance, where a p-value of $< 0.05$ indicates that the variances between groups are significantly different from each other. Homoscedasticity was visually inspected by plotting residuals versus fitted values in a scatterplot. Lastly, multicollinearity was investigated by examining correlations between predictors. Spearman's rank correlation was used to check the correlation between two categorical features or between categorical and numerical features, and Pearson's correlation was used between two numerical features. A correlation above 0.7 was considered high. 

Due to our study design that records repeated measures per participant, individual responses are not independent. To account for this within-participant correlation, we used mixed effects models with random intercepts for participants in our analyses. This way, within-participant variability is accounted for without estimating participant-specific effects.

\paragraph{Hypotheses}

To test for the first hypothesis, we utilized data collected in the first two blocks. Three mixed effects models were fit with the absolute difference in estimates as dependent variable (absolute difference in house price estimates collected in the first two blocks, and absolute difference in AI estimate collected in the second block). Smaller values in the absolute differences represent better performance. A boxcox transformation was applied to the dependent variables to ensure that normality assumptions are sufficiently fulfilled. Predictors were added to the model step by step, starting with treatment groups as the sole predictors, and then gradually adding potential demographic confounders (\textit{gender}, \textit{education}, \textit{age}), experience-related confounders (\textit{practical AI experience}, \textit{theoretical knowledge of AI}, \textit{frequency of work with AI}, \textit{recent house buying experience}), and study design factors (\textit{explanation view mode}). 

For each trial in the third block, a binary variable was recorded indicating whether participants chose to follow the AI or not. We assume that participants that trust the AI more will follow the AI's advice more often. As the outcome is binary, we fit a logistic regression model to estimate the probability of a user choosing the AI. A mixed effects model with random intercepts was fit to the data. As before, predictors were added in a step-wise manners. For models that did not reach convergence, a Bayesian equivalent was estimated.


\newpage

\subsection*{Data and code availability}

Due to privacy reasons we cannot make the data publicly available. The code that support the analysis and findings of this research is available at \url{https://github.com/tabearoeber/counterfactual_study_analysis}. 

\subsection*{Acknowledgements}

We extend our gratitude to Mahon Hughes for his assistance in submitting applications for ethical approval and data registration.

\subsection*{Supplementary information}

We include details on trained models, such as model coefficients and assumption checks in our supplementary material.

\subsection*{Declaration of interests}
The authors declare no competing interests.

\subsection*{Ethical declarations}

The study obtained ethical approval by the Research Governance and Integrity Team (RGIT) of Imperial College London as well as by the Economics and Business Ethics Committee (EBEC) of the University of Amsterdam.

\end{document}